\documentclass[journal,twoside]{IEEEtran}
\usepackage{cite}
\usepackage{graphicx}

\pdfoutput=1

\newcommand{\intensity}[2]{#1$\times$10$^{#2}$~\textrm{W/cm$^2$}}

\newcommand{\dist}[1]{#1~$\mu$m}

\newcommand{\ri}[2]{#1$\times$10$^{#2}$}

\newcommand{\loss}[1]{#1~dB/cm}

\newcommand{\speed}[1]{#1~$\mu$m/s}

\newcommand{\energy}[1]{#1~$\mu$J}

\newcommand{\obj}[2]{#1$\times$ (#2 NA)}

\begin{document}
\title{Investigation of ultrafast laser--photonic\\material interactions: challenges for\\directly written glass photonics}

\author{M.~Ams,
        G.~D.~Marshall,
        P.~Dekker,
        M.~Dubov,
        V.~K.~Mezentsev,
        I.~Bennion
        and M.~J.~Withford,~\IEEEmembership{Member,~OSA}%
\thanks{Manuscript received February 11, 2008; revised February 11, 2008.}%
\thanks{M.~Ams (mams@physics.mq.edu.au), G.~D.~Marshall, P.~Dekker and~M.~J.~Withford are with the MQ Photonics Research Centre, the Centre for Ultrahigh bandwidth Devices for Optical Systems (CUDOS) and the Department of Physics, Macquarie University, North Ryde, NSW 2109, Australia.}%
\thanks{M.~Dubov and I.~Bennion are with the Photonics Research Group, Electronic Engineering, Aston University, Aston Triangle B4 7ET Birmingham, United Kingdom.}}



\IEEEspecialpapernotice{(Invited Paper)}

\maketitle

\begin{abstract}
Currently, direct-write waveguide fabrication is probably the most widely studied application of femtosecond laser micromachining in transparent dielectrics. Devices such as buried waveguides, power splitters, couplers, gratings and optical amplifiers have all been demonstrated. Waveguide properties depend critically on the sample material properties and writing laser characteristics. In this paper we discuss the challenges facing researchers using the femtosecond laser direct-write technique with specific emphasis being placed on the suitability of fused silica and phosphate glass as device hosts for different applications.
\end{abstract}


\section{Introduction}
\IEEEPARstart{I}{n} 1996, it was shown that tightly focussed femtosecond infrared laser pulses can create a permanent refractive index modification inside bulk glass materials\cite{davis1996,glezer1996}. Although investigations into understanding the nature of this modification and the conditions that produce it are ongoing, it is widely accepted that the modification process is initiated by the rapid absorption of laser energy through nonlinear excitation mechanisms\cite{schaffer2001a}. The subsequent dissipation of this energy into the lattice causes the modification inside the glass. This result enables the direct-write fabrication of optical devices, active and passive, in a variety of dielectric optical materials including amorphous glasses, crystalline materials and optical polymers simply by moving the glass sample through the focus of a femtosecond laser beam. The material surrounding the focal volume remains largely unaffected by the writing beam passing through it, allowing structures to be written at arbitrary depths and in a three-dimensional fashion.

The femtosecond laser direct-write technique has been used to fabricate buried waveguides\cite{davis1996, homoelle1999, efimov2001}, power splitters\cite{nolte2003, liu2005, low2005}, couplers\cite{minoshima2002, watanabe2003, kowalevicz2005, suzuki2006}, gratings\cite{martinez2004, sudrie1999, kawamura2001, florea2003, takeshima2005, liu2007} and optical amplifiers\cite{sikorski2000, osellame2003, dellavalle2005, psaila2007}. These devices have been produced using (i) regeneratively amplified Ti:Sapphire laser systems that provide high pulse energies ($\mu$J-mJ) at kHz repetition rates, (ii) oscillator-only Ti:Sapphire systems with low energy (nJ) and high repetition rates (MHz),  (iii) high pulse energy (nJ-$\mu$J) ytterbium-doped fibre lasers at high repetition rates (100~kHz-MHz) as well as cavity dumped Yb:KYW laser oscillators. While all of the systems described above are effective at modifying transparent dielectrics significant differences exist between the mechanism underlying the modification and therefore also the strength of the modification, level of damage (if any) and most importantly in terms of waveguides whether the index change is positive or negative. Key parameters which affect the writing properties include the sample translation speed, focussed beam shape, beam polarisation, pulse energy, pulse repetition rate, wavelength and pulse duration. Other properties that dictate the type of material modification include, for example, bandgap energy, whether the sample is crystalline or amorphous, thermal characteristics and fracture strength. In this paper we review work in the area of direct-write femtosecond laser modification of photonic materials with an emphasis on fabricating waveguides devices in silica and doped phosphate glass using a high energy, low repetition rate Ti:sapphire laser amplifier and for comparison, a high repetition rate, low energy oscillator-only laser system.
\IEEEpubidadjcol 

\section{Materials}
The materials interaction processes at play within the laser focus are strongly dependant on both the material and the laser parameters, and it is common to observe both positive and negative changes in the material refractive index under different laser processing conditions or even within the same interaction region. Most studies into the fundamental physical processes that occur at the laser focus have been conducted in fused silica. In comparison with other optical materials, fused silica can be processed under a wide range of laser pulse frequencies, durations and energies, wavelengths and sample translation speeds. Furthermore, fused silica is easy to obtain in high purity forms by virtue of it being a popular UV optical material. Borosilicate glass has also been extensively studied. The most important property of borosilicate glasses is, however, the response of the glass to cumulative heating resulting from high ($>$500~kHz) pulse repetition frequency laser exposure. Borosilicate glasses (in contrast to fused silica) have been demonstrated to exhibit controlled growth of the heat affected zone centred at the laser focus within the material thus controlling the dimensions of the written optical waveguide\cite{eaton2005}. This facile control of the heat affected zone through judicious selection of the laser processing parameters is an example of how the combination of the correct material and laser processing parameters can be used to great effect in the creation of arbitrary waveguide designs.

Fused silica and borosilicate glasses provide an excellent platform in which to create passive optical devices. The solubility of active, rare-earth, ions in these glasses is low and despite the extensive use of rare-earth ion doped silica glasses in optical fibre devices, the relatively low gain-per-unit length value (0.3~dB/cm) of these materials makes it difficult to realise high-gain devices in a typical few centimetres long directly-written device. Consequently there has been a great deal of interest in phosphate glass hosts in which tens of percent by weight of rare-earth ions can be held in solution offering a higher gain-per-unit length value (4~dB/cm) without detrimental effects such as ion-clustering. Erbium and Ytterbium co-doped phosphate glass hosts have been successfully laser processed and used to create optical amplifiers and, with the addition of external reflectors, optically pumped waveguide-lasers (WGL)\cite{dellavalle2007}.

Apart from the passive and active glass materials typically used for directly-written devices, the femtosecond laser direct-write technique has also been applied to common crystalline materials such as LiNbO$_3$\cite{bookey2007}, YAG\cite{okhrimchuk2005, dubov2004}, LiF\cite{kawamura2004} and Ti:Sapphire\cite{apostolopoulos2004}. The dominant material change, when using a low repetition rate femtosecond laser, in most of these materials is a negative refractive index change, however, use of suppressed cladding arrangements or induced stress fields allowed waveguiding regions to be realised.

\section{Materials Interaction Processes}
\subsection{Nonlinear Excitation Mechanisms}
In non-metallic materials, the valence band is the highest occupied energy level where electrons are normally present at absolute zero, i.e. the lowest band of allowed states. Since electrons have a tendency to fill the lowest available energy states, the valence band is always nearly completely filled with electrons. An energy gap, $E_g$, separates the valence band from the conduction band; the lowest unoccupied energy level in the material. When  valence electrons gain enough energy, from a radiation field for example, they can leave the valence band to rise up to the conduction band and become free electrons. To achieve such a promotion, the radiation field's photon energy must exceed the bandgap energy, $E_g$. Typically, a single photon of visible light does not possess enough energy to exceed the bandgap energy of typical optical materials. In this case, nonlinear absorption processes are required to promote valence electrons to the conduction band. This can be accomplished through photoionisation and avalanche ionisation~\cite{schaffer2001a}. Photoionisation refers to the direct excitation of electrons by the laser field and can be broken down into two different regimes which are dependant on laser frequency and intensity: multiphoton ionisation (MPI) and tunnelling ionisation.

\subsubsection{Tunnelling Ionisation}
\begin{figure}[!t]
\centering
\includegraphics[width=2.5in]{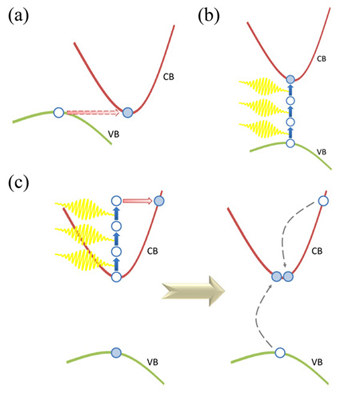}
\caption{Nonlinear photoionisation processes underlying femtosecond laser machining. (a) Tunnelling Ionisation, (b) Multiphoton Ionisation and (c) Avalanche Ionisation: Free carrier absorption followed by impact ionisation. VB - Valence Band, CB - Conduction Band.}
\label{ionisation}
\end{figure}

Upon radiation, the band structure of a dielectric can be distorted due to the presence of an electromagnetic (EM) field. This field suppresses the potential that binds a valence electron to its parent atom. If the laser field is strong enough, band to band transitions can occur whereby a bound electron tunnels out to become a free electron. This is illustrated in Fig.~\ref{ionisation}(a). Tunnelling ionisation is the dominant nonlinear ionisation regime for strong laser fields and low laser frequencies~\cite{schaffer2001a}.

\subsubsection{Multiphoton Ionisation (MPI)}
MPI occurs due to the simultaneous absorption of multiple photons by a single electron in the valence band. In order for an interband transition to occur, the total absorbed energy of all the $n$ photons that interact with the single electron must have an energy that exceeds the bandgap energy $E_g$, i.e. $n\hbar\omega \geq E_g$ where $\hbar$ is Planck's constant and $\omega$ the laser frequency. In fused silica, which has a bandgap energy of 9~eV, at least 6 photons are required to be absorbed by a single valence electron to drive an interband transition. MPI is illustrated in Fig.~\ref{ionisation}(b). MPI is typically associated with high laser frequencies (still below that required for linear absorption)~\cite{schaffer2001a}.

\subsubsection{Avalanche Ionisation}
An electron already in the conduction band can also sequentially absorb several laser photons until its energy exceeds the conduction band minimum by more than the bandgap energy $E_g$ (see Fig.~\ref{ionisation}(c)). This electron can then collisionally ionise (via impact ionisation) another electron from the valence band, resulting in two electrons in the conduction band's lowest available energy state~\cite{schaffer2001a, stuart1996}. This process will repeat as long as the laser field is present causing the conduction band electron density to increase exponentially. Kaiser \textit{et al}. showed that avalanche ionisation typically develops for pulse durations greater than 200~fs~\cite{kaiser2000}.

\subsection{Energy Transfer}
The ionisation process results in a transfer of energy from the radiation field to the dielectric's electrons creating a free electron gas. Eventually this deposited energy is redistributed over the various energy states of the system, i.e. the energy is then transferred from the electrons to the lattice via electron-phonon coupling. It is important to note that femtosecond nonlinear absorption occurs on a time scale that is short compared to the time scale of energy transfer within the system; the time it takes for the electrons to transfer their energy to the lattice. As a consequence the absorption and lattice heating processes can be decoupled and treated separately~\cite{stuart1996}. In essence, a femtosecond laser pulse produces a strong non-equilibrium condition in a material with electron temperatures much higher than lattice temperatures. In other words, at the end of the femtosecond laser pulse there are many hot electrons within a cold lattice. How a system reacts to these strong non-equilibrium conditions determines the process of energy relaxation and the types of structural changes that can be produced inside a material.

\section{Modification Regimes}
\begin{figure}[!t]
\centering
\includegraphics[width=3.4in]{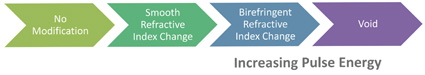}
\caption{Material modification regimes of (most) transparent glasses induced by the femtosecond laser direct-write technique.}
\label{modregimes}
\end{figure}

Three different types of material modification have been induced in the bulk of transparent materials using the femtosecond laser direct-write technique; a smooth isotropic refractive index change, a birefringent refractive index change and a void. Fig.~\ref{modregimes} illustrates these modification regimes for fused silica (and most other transparent glasses) induced by femtosecond laser pulses.

\subsection{Smooth Refractive Index Change}
\subsubsection{Thermal Model}
Although the explicit mechanism that may contribute to refractive index change is not known, it seems likely that energy deposited into the focal volume of a material by near-threshold femtosecond laser irradiation leads to local rapid heating and modification of a small volume of the glass at the focal spot~\cite{chan2001, chan2003a}. Because the thermal gradient achieved by this process is localised to the focal point, a very small fraction of the whole lattice, the glass subsequently cools very rapidly. In fused silica, the density increases if the glass is quenched from a high temperature~\cite{bruckner1970, bruckner1971, haken2000}, explaining the higher refractive index (typically on the order of $10^{-3}$) observed in femtosecond laser irradiated fused silica~\cite{davis1996, homoelle1999, will2002}. Similar results for other glasses~\cite{chan2003, nolte2004, bhardwaj2005, thomson2006, siegel2005} show either an increased or decreased\footnote{Note that even in glasses where irradiation leads to a refractive index decrease in the irradiated volume, there is often a refractive index increase just outside this volume, likely due to compression and densification of the material that surrounds the, now less dense, irradiated volume~\cite{nolte2004}.} refractive index change with increasing cooling rates confirming that the induced index change is related to a thermal process. However, a model of a thermal origin for the index change showed that low energy infrared oscillator exposure and high energy infrared amplifier exposure do not achieve the same temperatures during fabrication, yet they induce similar index changes~\cite{streltsov2002}. This result suggests that thermal heating is not the only mechanism that can lead to a smooth refractive index change.

\subsubsection{Colour Centre Model}
It has been suggested that the effect of radiation produces colour centres~\cite{davis1996, hirao1998, efimov2001, chan2003, chan2003a} in sufficient numbers and strength to alter the refractive index through a Kramers-Kronig mechanism~\cite{streltsov2002}. This theory has been a proposed mechanism for the index change produced by deep-UV excitation of Ge-doped silica fibres that result in fibre Bragg gratings~\cite{hill1997}. A high electron density, produced by the nonlinear absorption mechanisms outlined above, leads to a sufficient trapped species (colour centres) concentration in the exposed region resulting in different types of substrate defects being formed. Confocal fluorescence spectroscopy at 488~nm has been used by a number of different research groups to detect changes in the molecular structure within femtosecond laser irradiated regions~\cite{hirao1998, davis1996, chan2003a, reichman2007}. According to standard electron spin resonance (ESR) investigations of irradiated fused silica glass\footnote{Fused silica has been outlined here as an example and because it has been actively researched. Other defects associated with other glasses do exist and can be found in the literature, for example~\cite{efimov2001, chan2003}.}, a fluorescence peak at 630~nm due to non-bridging oxygen hole centre (NBOHC) defects is produced as well as a peak centred at 540~nm, characteristic of self-trapped exciton SiE' defects from small silicon nanoclusters. This direct evidence of colour centre formation in a femtosecond laser modified region may contribute to the refractive index changes also associated with femtosecond laser modification. These colour centres, however, do not produce the majority of the induced refractive index change because eliminating them by annealing (photobleaching) does not recover the original index~\cite{will2002, streltsov2002, sudrie2001}.

\subsubsection{Structural Change Model}
Poumellec~\textit{et al}.~\cite{poumellec2006} showed that densification and strain in the glass due to femtosecond laser radiation may also account for changes in the index of refraction. In order to detect changes in the types of network structures within a glass material, researchers used Raman spectroscopy~\cite{chan2003a, reichman2006, reichman2007}. Fused silica typically has large 5- and 6- fold ring structures dominant in its network~\cite{pasquarello1998, kubota2001}. However, scattering from a femtosecond laser modified region of fused silica resulted in Raman peaks centred at 490~cm$^{-1}$ and 605~cm$^{-1}$, which were attributed to the breathing modes of 4- and 3- membered ring structures in the silica network respectfully~\cite{galeener1984, pasquarello1998}. These low rank rings are a sign of elevated energy in the silica structure consistent with the nonlinear absorption mechanisms mentioned earlier. An increase in the 3- and 4- fold ring structures (and an associated decrease in the number of 5- and 6- fold ring structures) present in femtosecond laser modified regions leads to a decrease in the overall bond angle in the silica network and a densification of the glass~\cite{kubota2001, kawamura2001, efimov2001, chan2001, marcinkevicius2001a}. It has also been shown that the refractive index and the abundance of these low rank ring structures in femtosecond laser exposed regions increase in the same way. Furthermore, Hirao~\textit{et al.} examined the laser modified region by an atomic force microscope (AFM) and showed that a refractive index change is related to this densification process~\cite{hirao1998a}. In contrast to colour centre photobleaching, the Raman changes observed in the network structure remain permanent.

A common feature of laser induced densification is the stress that is produced in the surrounding unexposed medium in response to volume changes produced in the exposed region. These stresses manifest themselves as birefringence~\cite{homoelle1999, streltsov2002}. Assuming a uniform densification within the femtosecond laser modified region, the relative magnification of the induced density change can be calculated from the measured birefringence~\cite{homoelle1999, streltsov2002}. Such measurements have indicated that densification alone cannot account for the entire change in the index of refraction~\cite{homoelle1999, streltsov2002, allan1996}.

\subsubsection{Summary}
Smooth refractive index change induced by femtosecond laser radiation is likely due to a contribution of all the effects outlined above, i.e. colour centre formation, densification (structural change) and thermal treatment (melting) of the glass. Optical waveguide devices are fabricated in materials using design parameters that give rise to this regime of modification.

\subsection{Birefringent Refractive Index Change}
Under slightly different parameters, it has been shown that modified regions in fused silica using the direct-write technique contain nanoporous structures that are dependent on the polarisation of the femtosecond laser writing beam~\cite{shimotsuma2003, hnatovsky2006, poumellec2003}. These nanostructures were found to be self-ordered and periodic (with a size and period as low as 20~nm and 140~nm respectively) while being orientated in a direction perpendicular to the electric field vector of a linearly polarised femtosecond laser writing beam.

Using a scanning electron microscope (SEM) and selective chemical etching, researchers were able to show that the nanostructures consist of alternating regions of material with slightly increased density and slightly decreased density. This periodic varying material composition found in the irradiated volume gives rise to birefringent refractive index changes~\cite{sudrie1999, sudrie2001}. Furthermore, Auger electron spectroscopy of the same regions revealed that the concentration of oxygen varies across the irradiated area~\cite{hirao2005, shimotsuma2003}. These results indicated that the periodic nanostructures consist of periodically distributed oxygen deficient regions.

Two explanations for the formation of these nanostructures or `nanogratings' have been postulated. Shimotsuma~\textit{et al.} argue that the interference between the incident light field and the electric field of the bulk electron plasma wave, induced via nonlinear absorption, results in a periodic modulation of electron plasma concentration and permanent structural changes in the glass network~\cite{shimotsuma2003, hirao2005}. This theory would serve as the first direct evidence of interference between light and electron density waves. Hnatovsky~\textit{et al.} suggest the evolution of nanoplasmas into disc shaped structures due to high nonlinear ionisation creates the nanostructures~\cite{hnatovsky2006}. The observed nanostructures represent the smallest embedded structures created to date using light. Thus far, this induced `form birefringence' has been shown to only exist in fused silica.

\subsection{Void}
At extremely high intensities, the region of modification is characterised by material damage or void formation. Due to avalanche ionisation and continuous impact ionisation, a localised plasma is formed in the focal region~\cite{glezer1996, glezer1997}. As the temperature increases in the exposed region, the plasma causes a large charge separation resulting in high pressures. This charge separation is sufficient enough to cause a Coulomb explosion (microexplosion) generating a shock wave~\cite{ashkenasi2003, glezer1997, qiu1998}. Because this explosion or expansion occurs within the bulk of a material, the shock wave carries matter and energy away from the focal volume, compressing the surrounding material and leaving a rarified (less dense or hollow core) central region termed a void~\cite{schaffer2004, gorelik2003}. The contention that shock waves exist during femtosecond laser modification with high pulse energies is supported by the detection of acoustic or pressure waves originating from the focal point~\cite{horn2004, sakakura2007, sun2007}. Voids have been used in for the fabrication of optical memory devices~\cite{watanabe2000}, fibre Bragg gratings~\cite{marshall2006} and 2D waveguide arrays~\cite{mendez2007}.

\section{Experiment}
\subsection{Fabrication}
\begin{figure}[!t]
\centering
\includegraphics[width=3.4in]{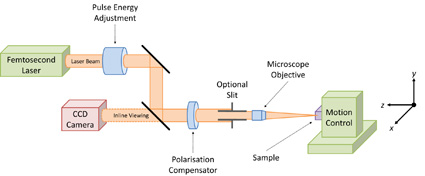}
\caption{Writing setup used to fabricate optical waveguide devices.}
\label{writesetup}
\end{figure}

In this paper, optical waveguide devices were manufactured using either a regeneratively amplified Ti:sapphire Spectra Physics Hurricane laser (pulse length 120~fs, repetition rate 1~kHz) or a Ti:sapphire Femtolaser XL oscillator (pulse length 60~fs, repetition rate 11~MHz) together with the setup shown in Fig.~\ref{writesetup}.
The 800~nm beam exiting the femtosecond laser passed through a computerised rotatable 1/2-wave plate and linear polariser setup allowing fine control of the pulse energy to be achieved. The femtosecond laser pulses were then focussed into the glass sample using a microscope objective. A variety of objectives with different numerical aperture (NA) and working distances were used so that the size and shape of the fabricated structures could be tailored to a certain degree. Typically, high NA objectives were used in conjunction with the XL oscillator system as a tight focus was required to reach an intensity in order to modify the glass substrate. Such a tight focus was not required when using the amplified Hurricane system. The microscope objectives we used are shown in Table~\ref{microobj}.
\begin{table}[!t]
  \renewcommand{\arraystretch}{1.3}
  \caption{Microscope Objectives.}
  \label{microobj}
  \centering
    \begin{tabular}{c|c|c|c|c}
    \hline
    Mag. & Type & Supplier & NA & Working \\
    & & & & Distance (WD) \\ \hline\hline
    20$\times$ & UMPlanFL & Olympus & 0.46 & 3.1~mm \\
    50$\times$ & UMPlanFL & Olympus & 0.8 & \dist{660} \\
    60$\times$ & 0.17~mm cover & Nikon & 0.85 & \dist{330} \\
    & slip corrected & & & \\
    \hline
    \end{tabular}
\end{table}
Before entering the microscope objective, the polarisation of the laser beam could be adjusted using a polarisation compensator (New Focus Model 5540). When the Hurricane laser was used, the physical shape of the laser pulses were modified by a horizontal slit aperture positioned before the microscope objective. The slit (which was orientated with its long dimension in the direction of sample translation) served to expand the laser focus in the direction normal to the laser beam propagation and sample translation. This enabled waveguides with circular symmetry to be written using a low magnification, long working distance objective\cite{ams2005, ho2004}. The Hurricane's amplified laser output could also be square-wave modulated in intensity using an external frequency source, thereby creating a waveguide structure formed by segments of exposed glass with a desired period. This technique was used, for example, to fabricate waveguide Bragg gratings\cite{marshall2007}. Pulse energies ranging from 0.005-10~$\mu$J, measured before focussing (and after passing through the slit if used), were used in the formation of optical waveguide devices. Using an air-bearing computer controlled XYZ stage (Aerotech), glass samples were scanned in a direction perpendicular to the direction of beam propagation, at speeds ranging between 25~$\mu$m/s and 10~mm/s.

\subsection{Materials \& Material Processing}
The two glass materials used in this paper were high quality grade fused silica from Schott AG (Lithosil Q0, $n_{800}\approx 1.454$) and a custom Er/Yb co-doped phosphate glass melt from Kigre Inc. (QX 2\%~wt Er, 4\%~wt Yb, $n_{800}\approx 1.52$ ). Glass samples were cut to size (diamond disc blade) and ground and polished (Logitech PM4) before device fabrication. A high grade optical-polished surface is typically required since defects in the surface through which the writing-laser is transmitted can cause waveguide-defects that contribute to propagation losses. After device fabrication, both the input and output faces of the device were ground and polished back by approximately 150~$\mu$m so that clean and uniform entry and exit points of the device could be accessed for characterisation. All the glass samples used were not thermally treated before or after fabrication.

\subsection{Characterisation}
\begin{figure}[!t]
\centering
\includegraphics[width=3.4in]{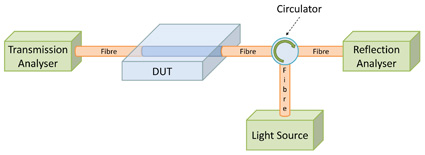}
\caption{Experimental setups used to take various transmission, reflection and propagation loss measurements. DUT - Device Under Test.}
\label{proploss}
\end{figure}

Typically, all waveguide devices were characterised in terms of their transmission and reflection data, near and far-field mode distributions, insertion, coupling, propagation and polarisation-dependent losses, induced refractive index contrasts and finally device gain. The experimental setup used to take transmission and reflection measurements from the device under test (DUT) is shown in Fig.~\ref{proploss}. %
Light sources including 635~nm, 976~nm, 980~nm and tunable C-band laser diodes were used to probe fabricated devices. Optical spectrum analysers (OSAs), power meters and CCD cameras were used to analyse device properties. Characterisation fibres were aligned to the DUT using 6-axis flexure stages. Index matching gel was used to reduce losses whenever optical fibres were used to either pump light into or collect light from the DUT.

A computational method\cite{mansour1996} was used to estimate the peak refractive index change between the bulk material and the waveguide structures. Transverse and end-on images of fabricated waveguide devices were taken in both reflection and transmission with Olympus differential interference contrast (DIC) microscopes.

The insertion loss (IL) of a fabricated device was taken to be the ratio of the measured transmitted powers with and without the DUT in the setup shown in Fig.~\ref{proploss} and included coupling, propagation and absorption losses. The coupling losses were estimated using the technique described in\cite{thomson2006} whilst absorption losses are material specific and can be measured using a spectrometer. The propagation loss was determined by taking the difference between the IL in reflection when the collecting fibre in Fig.~\ref{proploss} is replaced with a highly reflecting mirror aligned square to the device's output and the IL in transmission without the mirror.

The setup shown in Fig.~\ref{proploss} was slightly modified for active waveguide characterisation in that wavelength division multiplexers (WDMs) were inserted at the device's outputs so that both a signal source and pump source could co-propagate along the device in a bidirectional configuration.

\section{Design Considerations}
\subsection{Laser Repetition Rate}
Already mentioned was the fact that when using high repetition rate femtosecond laser systems hundreds of pulses accumulate to heat the focal volume constituting a point source of heat within the bulk of the material. Longer exposure of the material to this heat source gives rise to higher temperatures at focus resulting in a larger affected region~\cite{schaffer2003, eaton2005}. Due to symmetric thermal diffusion outside of the focal volume, a spherically shaped modified region is produced. In contrast, when using a low repetition rate femtosecond laser system the focal volume returns to room temperature before the arrival of the next pulse resulting in the same region of the material being heated and cooled many times by successive pulses. This repetitive type of machining means that the structural modification of the material is confined to the focal volume alone. If a low NA objective ($<$0.5 NA) is used in conjunction with a low repetition rate femtosecond laser, the focal volume (and hence the modified material volume) becomes asymmetric\cite{ams2005}. Beam shaping techniques, such as the slit method previously outlined, or multiple fabrication raster scans of the writing beam\cite{thomson2006, nasu2005, liu2004a} need to be employed to match the Rayleigh length with the focal spot diameter (confocal parameter).

\subsection{Spherical Aberration}
\begin{figure}[!t]
\centering
\includegraphics[width=3in]{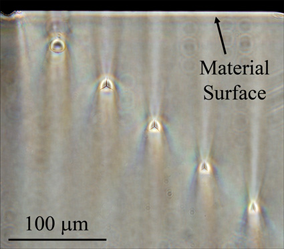}
\caption{Effects of spherical aberration on sub-surface focussing. Waveguide cross-sections become more asymmetric with deeper material penetration. A pulse energy of \energy{1.6} and a translation speed of \speed{25} were used to fabricate these waveguides.}
\label{sphaber}
\end{figure}

Most fabricators of waveguides use microscope objectives given that they are well corrected with high numerical apertures. However, directly written waveguides are created below the surface of a substrate and accordingly it is important to consider the effect of spherical aberration that sub-surface focusing causes. Fig.~\ref{sphaber} provides an example of the effect of spherical aberration on sub-surface focussing using a low repetition rate femtosecond laser.
An objective designed for surface observations was used to create an array of waveguides at decreasing depths. The waveguide cross-sectional shape changes from an aberrated and triangular one deep under the surface (aspect ratio of 0.64) to a more circular section closer to the surface of the glass (aspect ratio of 1.1). Spherical aberration and its effects on the waveguide cross section can be controlled using objectives that are corrected for focusing through a fixed depth of material (for example a cover-slip corrected objective) however this limits the 3-dimensional capabilities of the writing platform. A more suitable solution is to use oil-immersion focusing objectives that are not sensitive to the depth of focus in the material since all optical path lengths to the focus remain constant\footnote{A suitable readjustment of the slit width (if the slit method is used) can also reduce the effects spherical aberration.}.

\section{Results \& Discussion}
To date our research program has focused on the development of processing methodologies enabling the fabrication of the key building blocks of photonic circuitry, namely passive and active devices including low-loss waveguides, splitters, gratings, amplifiers and lasers for use in optical telecommunication systems.

\subsection{Passive Devices}
In most optical materials the limit to the maximum refractive index change that can be induced using the direct-write technique is determined in-part by the maximum intensity of laser exposure that the material can tolerate without suffering destructive damage. At the other extreme, an initial change in the index of refraction occurs when the laser intensity at focus reaches a level to initiate the onset of nonlinear absorption processes in the material. In fused silica, a refractive index change can first be seen with a writing intensity of \intensity{2}{13} at 100~fs, 1~kHz. The threshold for damage in fused silica occurs at laser intensities around \intensity{1}{16}. Waveguides written with intensities greater than this value develop void-like inclusions and the propagation losses in the waveguides rapidly increase to unacceptable levels with writing intensity. In phosphate glass hosts, a refractive index change begins to occur at intensities corresponding to \intensity{6.5}{13} at 100~fs, 1~kHz and \intensity{1}{13} at 60~fs, 11~MHz respectively. At intensities above \intensity{2.5}{14} at 100~fs, 1~kHz and \intensity{2.6}{13} at 60~fs, 11~MHz\cite{graf2007}, waveguide structures contain a significant number of voids. It can be seen from these values that the fabrication window for creating refractive index changes in bulk fused silica is much wider than the corresponding window for phosphate glass.

\begin{figure}[!t]
\centering
\includegraphics[width=3.4in]{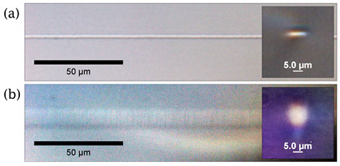}
\caption{Top-view DIC microscope images of waveguides fabricated with an intensity at focus of \intensity{2.8}{14} and a translation speed of \speed{25} in fused silica (a) without a slit and (b) with a 500~$\mu m$ slit positioned before the focussing objective. The insets show end-on white light transmission images of the respective waveguides.}
\label{slitdicfs}
\end{figure}
\begin{figure}[!t]
\centering
\includegraphics[width=3in]{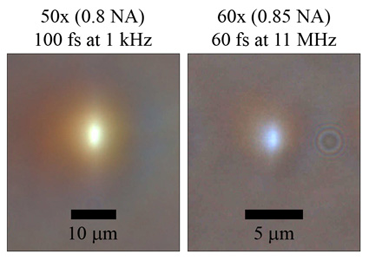}
\caption{End-on white light transmission micrographs of waveguides written in phosphate glass using different writing conditions: (Left) pulse energy -- \energy{0.25}, translation speed -- \speed{25}, slit width -- \dist{900} and (Right) pulse energy -- 19~nJ, translation speed -- 1~mm/s.}
\label{smallwg}
\end{figure}

%
Fig.~\ref{slitdicfs}(a) shows the top-view of a linear structure and its end-on cross-section fabricated with the low repetition rate system in fused silica without using any beam shaping techniques. Because of its elliptical cross-section, a guided circular mode could not be sustained along this structure. The waveguide shown in Fig.~\ref{slitdicfs}(b) was fabricated in the same sample using the same focussing objective and translation speed albeit with a \dist{500} slit aperture positioned before the focussing objective aligned parallel to the direction of sample translation. To generate the same intensity at focus (\intensity{2.8}{14}), as was used to create the waveguide shown in Fig.~\ref{slitdicfs}(a), the pulse energy was adjusted after passing through the slit. White light transmitted through the core of the waveguide fabricated using the slit method is clearly shown in the inset to have a circular diameter less than \dist{13}. The induced refractive index change of this waveguide was estimated to be \ri{5.2}{-4}. Typically, the propagation loss of fused silica waveguides at 1550~nm were measured to be \loss{0.83}. A similar study was conducted in phosphate glass at an intensity of \intensity{2.1}{14} revealing waveguides that guide a circular mode can be fabricated with propagation losses as low as \loss{0.39}\cite{ams2005}. As stated before, due to cumulative heating effects, the cross-section of waveguides written using a high repetition rate femtosecond laser oscillator are intrinsically circularly symmetric. The end-on white light transmission DIC image of such a waveguide, written in phosphate glass, is shown in Fig.~\ref{smallwg}.

\begin{figure}[!t]
\centering
\includegraphics[width=3in]{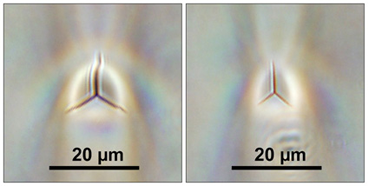}
\caption{End facet cracks from (Left) a symmetric waveguide written at a shallow depth and (Right) an asymmetric waveguide written more deeply. Image modified from Fig.~\ref{sphaber}.}
\label{endcrack}
\end{figure}

Whilst characterising the waveguides written in phosphate glass (outlined above), an additional loss mechanism was found after wet polishing due to cracking of the bulk material at device end facets. This phenomenon is thought to be caused by local stress relief as it is specific to the first 1-2~$\mu$m of the waveguide length. These cracks may be eliminated by fabricating waveguides at intensities below \intensity{1.9}{14} or by dry polishing the material. However, these stress cracks have been noted as revealing information regarding asymmetries in the waveguide form whilst also offering useful insights into the spatial nature of the embedded stress field. With increasing spherical aberration caused by sub-surface waveguide writing without a corrected objective, the end facet cracks are observed to increase in their deviation from a symmetric three radial 120$^{\circ}$ separated fracture. Fig.~\ref{endcrack} shows an excerpt from Fig.~\ref{sphaber} in which the end facet cracks from a circular waveguide written at a shallow depth are compared alongside a more deeply written and asymmetric waveguide. It could be argued that the asymmetric nature of the stress field for the deeper written waveguide will introduce stress related birefringence not observed in the shallow written waveguide.

\begin{figure}[!t]
\centering
\includegraphics[width=3in]{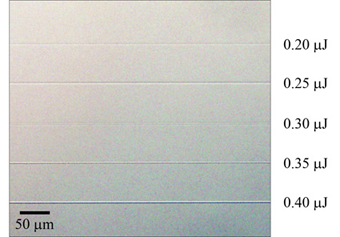}
\caption{Linear structures written with a translation speed of \speed{25} in phosphate glass using the \obj{50}{0.8} microscope objective and various pulse energies measured after a \dist{900} slit. The top two structures guide light indicating a positive index change. The bottom two structures do not guide light representing a negative index change. The structure in the middle shows the transition between a positive and negative index change.}
\label{phosneg}
\end{figure}

Phosphate glass also displayed an interesting property in that not just the magnitude but also the sign of the net refractive index change induced by the writing laser is a function of pulse energy. This effect is observed in high numerical aperture ($\geq$0.5~NA) focusing arrangements with the transition between positive and negative refractive index change occurring at intensities of approximately \intensity{3.6}{14} at 100~fs, 1~kHz and \intensity{3.1}{13} at 60~fs, 11~MHz. Increasing the laser intensity in these high numerical aperture focusing arrangements creates a greater magnitude of negative refractive index change. Fig.~\ref{phosneg} clearly shows this transition where positive index waveguides appear in different contrast to negative index structures when viewed using DIC microscopy techniques.
This manner of change was observed in both high repetition rate (11~MHz) and low repetition rate (1~kHz) waveguide writing configurations, using a number of microscope objectives, which rely on different photoionisation mechanisms to produce the refractive index modification. This result effectively indicates that the range of writing intensities available for producing positive index waveguide devices in phosphate glass is reduced to \intensity{6.5}{13}-\intensity{3.6}{14} at 100~fs, 1~kHz and \intensity{2.4}{13}-\intensity{3.1}{13} at 60~fs, 11~MHz.

We also compared and contrasted the optical transmission properties of straight and curved waveguides written with linearly and circularly polarised light (at 100~fs, 1~kHz) in fused silica glass, and showed an increase in transmission through waveguides written using circularly polarised light~\cite{ams2006}. This increase in light transmission is still under investigation but may be explained by a modification of the periodic aligned nanostructures that accompany devices fabricated with linearly polarised radiation~\cite{shimotsuma2003, hnatovsky2006}. Waveguides fabricated in fused silica using the slit method and a circularly polarised writing beam possess a propagation loss of approximately 0.83~dB/cm. The use of a circularly polarised writing beam has also been found beneficial in fabricating low-loss waveguides in LiNbO$_3$ samples\cite{nejadmalayeri2006}. 

It has been shown that the refractive index contrast of a modified region can be increased by overwriting a waveguide with more than one pass of a low repetition rate femtosecond laser beam in a multiple fabrication scan fashion~\cite{davis1996, hirao1998a, low2005}. We conducted a similar study and found that waveguides written in fused silica with 8 multiple passes exhibit an unsaturated propagation loss of approximately 0.36~dB/cm. The diameter of these waveguides did not change with increasing laser scans. This result is consistent with other reports that attribute increased refractive index changes with the number of fabrication passes\cite{hirao1998a}. Another possible explanation may be that waveguides become more `smooth' due to consecutive laser scans correcting waveguide imperfections produced by earlier scans. In the case of waveguides fabricated in phosphate glass, it was found that after 7 multiple passes of the writing beam, the index change became negative. We also found that as the translation speed decreases, the propagation loss also decreases. Unlike the multiple pass scheme, however, which showed that the propagation loss decreased linearly with the number of passes, the propagation loss associated with a reduction in translation speed seemed to saturate beyond an on target energy density of 0.06~$\mathrm{\mu J/\mu m^3}$, a value that would correspond in energy deposition to a waveguide being written with 6 multiple passes.

Typically, we found that all waveguide devices fabricated in fused silica remained stable up to a temperature of 600$^\circ$C. Devices fabricated in phosphate glass, however, were only stable up to a temperature of 350$^\circ$C. Beyond this value, the positive index change associated with the fabricated devices inverted and became negative. Investigations into the thermal characteristics of waveguide devices written in phosphate glass are ongoing. These results show that there are additional complications with phosphate glass whilst also underlining the importance of matching the correct host material to the target application.

\subsection{Active Devices}
Earlier is was pointed out that doped silica glasses have a low gain-per-unit length value (\loss{0.3}) which is comparable in size to the typical propagation losses ($\approx$\loss{0.2}) associated with femtosecond laser written waveguides fabricated in such glass materials\cite{hirao1998a}. Waveguides written in doped silicate glasses possess similar propagation losses, however, the gain-per-unit length of these waveguides can reach values near \loss{2}\cite{psaila2007}. There have been several reports of C-band amplifying waveguide devices created in Erbium/Ytterbium co-doped phosphate glass hosts\cite{osellame2002, osellame2003, taccheo2004, dellavalle2005}. These devices typically exhibit between \loss{2-4} of internal gain and less than \loss{0.5} of propagation loss, demonstrating that doped phosphate glass hosts may be more suited to active device fabrication than silica based materials.

\begin{figure}[!t]
\centering
\includegraphics[width=3in]{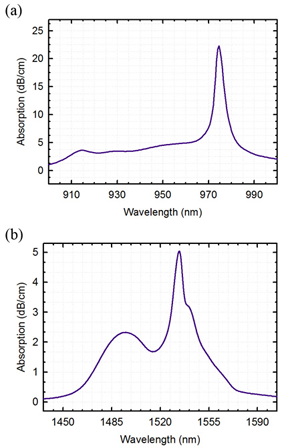}
\caption{Absorption spectra of Kigre co-doped QX glass. (a) Absorption due to the Yb$^{3+}$ ions (pump wavelength) and (b) Absorption due to the Er$^{3+}$ ions (signal wavelength).}
\label{erybspectra}
\end{figure}

\begin{figure}[!t]
\centering
\includegraphics[width=3in]{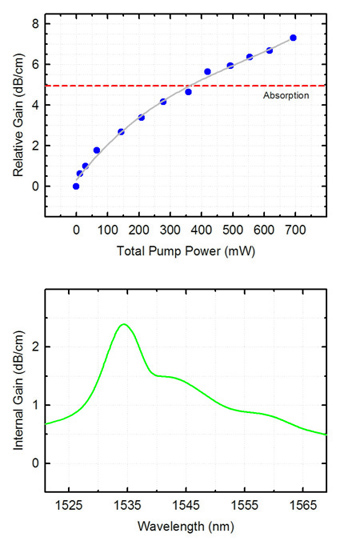}
\caption{(Top) Signal enhancement at 1534.6~nm versus total pump power. (Bottom) Internal gain of a waveguide amplifier fabricated in Er/Yb co-doped phosphate glass.}
\label{gain}
\end{figure}

Example absorption spectra of the Ytterbium and Erbium ions in a commonly employed phosphate glass host are shown in Fig.~\ref{erybspectra}(a) and (b) respectively. The Ytterbium absorption spectrum is characterised by a single dominant peak at 975~nm due to the $^2 F_{7/2}$ to $^2 F_{5/2}$ transition of the Yb$^{3+}$ ion which is used to optically pump the waveguide device. The Er$^{3+}$ absorption spectrum displays two broad absorption curves that are the result of the many host-field Stark split $^{4}I_{13/2}$ to $^{4}I_{15/2}$ transitions.
Linear waveguides were written in an Er/Yb co-doped phosphate glass sample with an intensity of \intensity{1.9}{14} and translation speed of \speed{25}. By supplying the maximum amount of available pump power to these waveguides, an internal gain (at approximately 1534~nm) of \loss{2.7} was obtained and optical amplification was shown to exist over the entire C-band (see Fig.~\ref{gain}). This result compares well with previous reports in the literature~\cite{osellame2006, thomson2006}.
There is still also potential for improvement of the internal gain figure. By optimising the amplifier's physical length and finding the optimal rare-earth ion doping concentrations, the overall internal gain is expected to increase.

Because the waveguide devices fabricated in the co-doped phosphate glass sample exhibit an internal gain, they can be used to create a laser oscillator. A waveguide resonator is formed by distributing the optical feedback over the entire length of the waveguide amplifier\cite{kogelnik1971, kringlebotn1994}. Several experimental techniques have been reported that enable the realisation of Bragg grating structures inside femtosecond laser written waveguides to create such a cavity\cite{marshall2006, zhang2007a}. By square-wave modulating the low repetition rate femtosecond laser's output, a first order waveguide Bragg grating (WBG) was fabricated in Er/Yb co-doped phosphate glass\cite{marshall2007}. Using a combined bidirectional pump power of 710~mW, a waveguide variant of a distributed feedback (DFB) laser was demonstrated using an external point source of heat to create the required $\pi/2$ phase shift mid-grating\cite{marshall2008}. The total output power of this laser measured 0.37~mW and had a linewidth $<$4~pm. Although it is known that the WBG structure contributes to an increase in the propagation loss of the amplifier device\cite{zhang2007a}, clearly the WBG is of a high enough quality that the internal gain in the system still exceeds this increase.

\section{Conclusion}
We reviewed the femtosecond laser direct-write technique as a technology capable of producing optical waveguide devices inside bulk transparent materials without the need for lithography, etching, a controlled environment or much sample preparation. A number of investigations into the challenges facing researchers using the femtosecond laser direct-write technique were undertaken. Most importantly, it was found that specific consideration of the pulse repetition rate and energy, writing beam polarisation, sample translation speed, number of fabrication scans, spherical aberration, polishing techniques and material preparation must be taken into account in order to fabricate low-loss positive index guiding waveguide devices in a specific type of glass. Our results highlight the complexities associated with the application of the femtosecond laser direct-write technique to phosphate glass hosts. In particular, phosphate glass, compared to fused silica, has tighter fabrication constraints with respect to pulse energy, wet polishing and thermal treatment. Nonetheless, the devices fabricated in both glass types outlined in this paper raise the prospect of creating optical devices for the use in aiding all-optical access communication networks.

\section*{Acknowledgment}
\addcontentsline{toc}{section}{Acknowledgment}
This work was produced with the assistance of the Australian Research Council under the ARC Centres of Excellence \& LIEF programs. Graham Marshall would like to thank the Australian Academy of Science for their financial assistance through the Scientific Visits to Europe scheme. M.~Dubov, V.~K.~Mezentsev \& I.~Bennion also acknowledge the financial support of the Engineering \& Physical Sciences Research Council in carrying out this work.
\ifCLASSOPTIONcaptionsoff
  \newpage
\fi

\end{document}